# Relaxation damping in oscillating contacts

M. Popov[1,2,*], V.L. Popov[1,2,3] and R. Pohrt[1]

[1]Berlin University of Technology, 10623 Berlin, Germany
[2] National Research Tomsk State University, 634050 Tomsk, Russia
[3]National Research Tomsk Polytechnic University, 634050 Tomsk, Russia

If a contact of two purely elastic bodies with no sliding (infinite coefficient of friction) is subjected to superimposed oscillations in the normal and tangential directions, then a specific damping appears, that is not dependent on friction or dissipation in the material. We call this effect "relaxation damping". The rate of energy dissipation due to relaxation damping is calculated in a closed analytic form for arbitrary axially-symmetric contacts. In the case of equal frequency of normal and tangential oscillations, the dissipated energy per cycle is proportional to the square of the amplitude of tangential oscillation and to the absolute value of the amplitude of normal oscillation, and is dependent on the phase shift between both oscillations. In the case of low frequency tangential motion with superimposed high frequency normal oscillations, the system acts as a tunable linear damper. Generalization of the results for macroscopically planar, randomly rough surfaces is discussed.

It is well known that oscillating tangential contacts exhibit frictional damping due to slip in parts of the contact. Solutions for this behavior in the case of spherical surfaces were given by Mindlin et al. [1] in 1952. This contact damping plays an important role in numerous applications in structural mechanics [2], tribology [3] and materials science [4]. Since this damping arises due to partial slip in the contact of bodies with curved surfaces, when the coefficient of friction tends towards infinity, slip disappears, frictional losses are eliminated, and the oscillation damping becomes zero [1]. However, when a contact oscillates in *both normal and tangential* directions, there is another, purely frictional loss mode that we refer to as "relaxation damping". To our knowledge this phenomenon has not yet been discussed in the literature. Damping due to a combination of normal and tangential oscillations has been studied recently by Davies et al. [5] for smooth two-dimensional profiles and by Putignano et. al. [6] for rough surfaces. However, the fact that dissipation exists even in the limiting case of an infinite coefficient of friction, when relative frictional movement of contacting bodies does not occur, went unnoticed. This effect is an example of purely "non-dissipative" damping, somewhat like the Landau damping in a collisionless plasma [7].

In its essence the proposed loss mechanism is similar to a spring that is deflected and abruptly released, converting the stored energy into elastic waves that are eventually dissipated. If we consider a body that is pressed into a plane, then moved tangentially (with "stick" conditions in the contact), and finally lifted in the normal direction, the accumulated shear energy will eventually be lost even if there is no slip in the contact area and the material is purely elastic. Thus, an apparently non-dissipative system shows dissipation. The same will also happen in contacts that oscillate normally and tangentially at the same time, even if the motion is very slow (quasi-static.) At first glance it seems contradictory that a slowly moving, non-dissipative system shows dissipation. The physical reason for this dissipation is the infinite stress concentration at the borders of a tangential contact. Due to the stress concentration, infinitely rapid movements occur in the material even in the case of quasi-static macroscopic movement of the contacting bodies, similar to the dissipation from elastic instabilities in the Prandtl-Tomlinson-model [8], [9],[10]. The physical nature of relaxation damping can be understood and analyzed very simply in the framework of the method of dimensionality reduction (MDR). For small oscillation amplitudes, the dissipation rate can be calculated analytically.

Consider a contact between two axially-symmetric elastic bodies with moduli of elasticity of $E_1$ and $E_2$, Poisson's numbers of $\nu_1$ and $\nu_2$, and shear moduli of $G_1$ and $G_2$, accordingly. We denote the difference between the profiles of the bodies as $\tilde{z} = f(r)$, where $\tilde{z}$ is the coordinate normal to the contact plane, and $r$ the in-plane polar radius. The profiles are brought into contact and are subjected to a superposition of normal and tangential oscillation with small amplitudes. This contact problem can be reduced to the contact of a rigid profile $\tilde{z} = f(r)$ with an elastic half-space, Fig. 1a.

In our analysis we use the method of dimensionality reduction (MDR) [11]. MDR is based on the solutions for the normal contact by Galin [12] and Sneddon [13] as well as their extensions for tangential contacts by Cattaneo [14], Mindlin [15], Jäger [16] und Ciavarella [17]. In the framework of the MDR, two preliminary steps are performed [11]: First, the three-dimensional elastic half-space is replaced by a one-dimensional linearly elastic foundation consisting of an array of independent springs, with a sufficiently small separation distance $\Delta x$ and normal and tangential stiffness $\Delta k_z$ and $\Delta k_x$ defined according to the rules

$$\Delta k_z = E^* \Delta x \quad \text{with} \quad \frac{1}{E^*} = \frac{1-\nu_1^2}{E_1} + \frac{1-\nu_2^2}{E_2}\ , \qquad (1)$$

$$\Delta k_x = G^* \Delta x \quad \text{with} \quad \frac{1}{G^*} = \frac{2-\nu_1}{4G_1} + \frac{2-\nu_2}{4G_2}\ . \qquad (2)$$

In the second step, the three-dimensional profile $z = f(r)$ is transformed into a one-dimensional profile according to

$$g(x) = |x| \int_0^{|x|} \frac{f'(r)}{\sqrt{x^2 - r^2}} dr\ . \qquad (3)$$

If the MDR-transformed profile $g(x)$ is indented into the elastic foundation and is moved normally and tangentially according to an arbitrary law, the force-displacement relations of the one-dimensional system will exactly reproduce those of the initial three-dimensional contact problem (proofs have been done in [18] and [11].) The MDR solution is as accurate as the solutions of Cattaneo [14] and Mindlin [1]: the solution contains an inaccuracy, which has been shown to be generally quite small [19]. From the correctness of the force-displacement relations, it follows that the work and the dissipated energy will be reproduced correctly as well.

In the following, we consider, without loss of generality, a rigid conical indenter

$$z = f(r) = r \tan\theta \qquad (4)$$

in contact with half-space, Fig. 1a.

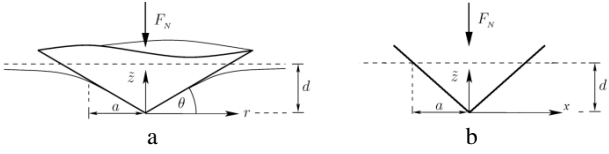

**Fig. 1:** (a) Contact of a cone with a half-space and (b) the corresponding MDR-transformed one-dimensional profile.

The one-dimensional MDR image of the conical profile (4), according to (3), is

$$g(x) = |x| \frac{\pi}{2} \tan\theta = c|x|, \tag{5}$$

where $c = (\pi/2)\tan\theta$ is the slope of the one-dimensional equivalent profile, Fig. 1b. The generalization for an arbitrary axis-symmetrical shape can be made very easily: if the amplitude of normal oscillation is sufficiently small compared to the indentation depth of the indenter, the shape of the edge of the contact will always be approximately linear. In this case, all axially-symmetric indenters will behave like conical indenters and the slope $c$ at the edge of the contact of the one-dimensional MDR-transformed profile will be the only shape-related parameter. For example, for a parabolic indenter $\tilde{z} = r^2/(2R)$, the MDR-transformed profile is $\tilde{z} = g(x) = x^2/R$ and the edge slope is $c = 2a/R$ where $a$ is the contact radius. The parameter $c$ can also be represented in a universal form that does not depend on the profile shape: The incremental contact stiffness is known to be equal to $\partial F_N/\partial d = 2aE^*$ [20]. Deriving this equation once more gives $\partial^2 F_N/\partial d^2 = 2E^* \partial a/\partial d = 2E^*/c$. Thus, the slope of the MDR-transformed profile can be calculated as

$$\frac{1}{c} = \frac{1}{2E^*} \frac{\partial^2 F_N}{\partial d^2}. \tag{6}$$

In the following, we consider energy dissipation in two cases: (a) oscillations in the normal and tangential direction with equal frequencies, (b) oscillation in the normal direction with much higher frequency than in the tangential direction.

(a) <u>Normal and tangential oscillations with equal frequencies.</u> Let the profile oscillate harmonically with a normal amplitude $u_z^{(0)}$, a tangential amplitude $u_x^{(0)}$ and a phase difference $\varphi_0$. To study the effect of relaxation damping in the pure, we assume an infinite friction coefficient between both bodies. Since the springs of elastic foundation in the MDR model are independent, it is sufficient to analyze the energy dissipation of a single spring (Fig. 2), and then to summarize over all springs which come into contact during an oscillation cycle. Consider a point of the rigid indenter with initial distance $z^{(0)}$ from the "surface". Its coordinates during the oscillatory motion can be written as $z(t) = -z^{(0)} + u_z^{(0)} \cos\omega t$ and $x(t) = x^{(0)} + u_x^{(0)} \cos(\omega t + \varphi_0)$. If $|u_z^{(0)}| > z^{(0)}$, the point of the rigid surface will come into contact with one of the springs of the elastic foundation in point $x_1$ and will drag it along to point $x_2$, where contact is lost and the spring relaxes over the distance $s = x_2 - x_1$. The coordinates $x_1$ and $x_2$ are determined by setting $z = 0$. After simple calculations we get

$$s = x_2 - x_1 = 2u_x^{(0)} \sqrt{1 - \left(z^{(0)}/u_z^{(0)}\right)^2} \sin\varphi_0. \tag{7}$$

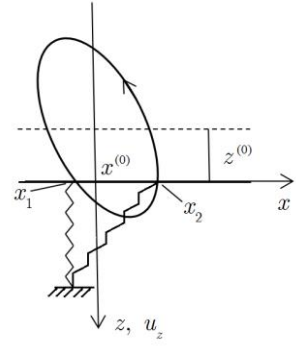

**Fig. 2:** A point of the rigid surface with the initial coordinate $z = -z^{(0)}$ oscillates around this position. It comes into contact with a spring in point $x_1$ and loses contact in point $x_2$.

The energy dissipated by a single spring during one cycle is equal to the energy stored in the stressed spring at the time of its release:

$$\Delta W = \frac{1}{2} \Delta k_x s^2 = \frac{1}{2} G^* s^2 \Delta x. \tag{8}$$

Energy dissipation occurs only if the point of the surface was in contact with the substrate during only a part of the cycle. This is the case for all points which satisfy the condition

$$-|u_z^{(0)}| < z^{(0)} < |u_z^{(0)}|. \tag{9}$$

Substituting $\Delta x = \Delta z^{(0)}/c$ in (8) and integrating over the interval (9), we obtain the total dissipated energy per cycle:

$$W = 2\frac{1}{2}\frac{G^*}{c} \int\limits_{-|u_z^{(0)}|}^{|u_z^{(0)}|} s^2 dz^{(0)}. \tag{10}$$

The factor "2" takes into account that there are two symmetric regions on both sides of the contact giving equal contributions to dissipation (this complete symmetry is only valid in the standard half-space-approximation used in this Letter). Substitution of (7) into (10) and evaluation of the integral finally gives

$$W = \frac{16}{3}\frac{G^*}{c} u_x^{(0)2} |u_z^{(0)}| \sin^2\varphi_0 \tag{11}$$

or in the shape invariant form, using (6),

$$W = \frac{8}{3}\frac{G^*}{E^*}\frac{\partial^2 F_N}{\partial d^2} u_x^{(0)2} |u_z^{(0)}| \sin^2\varphi_0 \tag{12}$$

(b) <u>Low frequency tangential motion with superimposed high frequency oscillation.</u> During one cycle of the normal oscillation, one can assume that the body is moving tangentially with a constant velocity $v_x^{(0)}$, $x = x^{(0)} + v_x^{(0)}t$, while the $z$-coordinate is still defined by $z = -z^{(0)} + u_z^{(0)}\cos\omega t$. The times at which a spring is coming into contact with the indenter ($t_1$) and is released ($t_2$) are given by the condition $z = 0$, from which it follows $t_{1,2} = \mp(1/\omega)\arccos\left(z^{(0)}/u_z^{(0)}\right)$. For the distance $s$, we get the following result:

$$s = x_2 - x_1 = 2\frac{v_x^{(0)}}{\omega}\arccos\left(z^{(0)}/u_z^{(0)}\right). \tag{13}$$

Substituting into (10) and evaluating of the integral, we get the energy dissipated per cycle:

$$W = 4\left(\pi^2 - 4\right)\frac{G^*}{c}|u_z^{(0)}|\frac{v_x^{(0)2}}{\omega^2} \tag{14}$$

or in the shape invariant form

$$W = 2\left(\pi^2 - 4\right)\frac{G^*}{E^*}\frac{\partial^2 F_N}{\partial d^2}|u_z^{(0)}|\frac{v_x^{(0)2}}{\omega^2}. \tag{15}$$

The dissipation power (energy dissipated per unit time), $P$, can now be simply evaluated by dividing (14) by the period of one oscillation, $2\pi/\omega$: $P = 2\left(\pi^2 - 4\right)G^*\left|u_z^{(0)}\right|v_x^{(0)2}/(\pi c\omega)$.

The dissipation power is proportional to the square of the velocity. Thus, the system acts as a simple damper with the damping constant $\gamma = 2\left(\pi^2 - 4\right)G^*\left|u_z^{(0)}\right|/(\pi c\omega)$, which can be tuned by changing either the amplitude or frequency of normal oscillations.

We would like to stress that in spite of the fact that the relaxation losses (11)-(12) and (14)-(15) have been derived in a one-dimensional model, they represent, due to the MDR theorems, the exact three-dimensional results for axis-symmetric profiles. In the shape invariant form (12) and (15) they are even applicable to multi-contact systems with independent contacts, as e.g. represented by the Greenwood and Williamson model of contact of rough surfaces [21]. This follows directly from the linearity of the energy losses with respect to the normal force. The shape invariance of the results (12) and (15) suggests that these may even be exact relations applicable to any three-dimensional contact topography. In order to verify this hypothesis, we carried out a series of three-dimensional, full-Cerruti-type numerical simulations of oscillating contacts using the methods described in detail in [22] and [23].

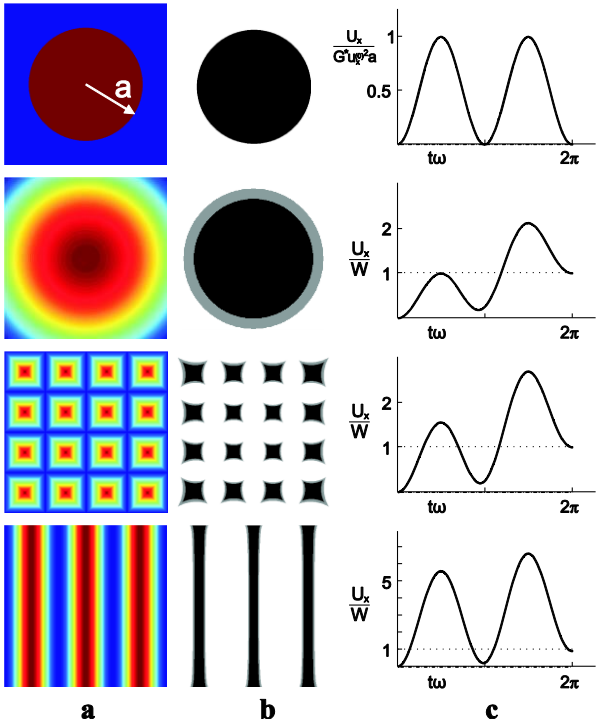

**Fig. 3** (a) Various surface profiles used to validate Eq. (12) by direct three-dimensional simulations of oscillating contact: a sharp-edged cylindrical profile; a parabolic surface; an arrangement of 16 pyramid indenters; a series of elongate sinusoidal profiles. (b) The contact configurations for the corresponding profiles. The minimum contact regions of a complete cycle are colored in black and the additional regions at maximum contact in gray. (c) Time plots of the work done by external forces in the x-direction on the system over one period of oscillation, normalized by the prediction $W$ according to eq. (12). In the first example, the contact area is not changed in the cycle so no dissipation takes place. In the other cases, the curves reach unity after one cycle, thereby confirming the validity of eq. 12. In all studied cases, the direct simulations reproduce the analytical result with an error not exceeding 5%.

The essential findings related to these simulations are summarized in Fig. 3: For 4 different surface topographies

(left column), the minimum and maximum contact areas are shown (middle column) as well as the time plots of the work done by the external force in the $x$-direction (right column). The total work done during one period (values reached at $\omega t$) is the dissipated energy. The horizontal dotted line shows the unity-normalization according to Eq. (12). One can see that the three-dimensional results coincide with the analytical prediction not only for axis-symmetrical profiles, but also for profiles having an "arbitrary" different form. We thus can conclude that Eq. (12) can be universally applied to contacts of arbitrarily shaped bodies. The same will be valid of course for Eq. (15).

Let us apply Eq. (15) to an important class of nominally flat rough surfaces (surfaces having a long wavelength cut-off of the power spectrum of roughness). For such surfaces, the relation between the normal force and the indentation depth is known to be $F_N \propto \exp(-d/l)$ [24], [11], where $l$ is the order of magnitude of the rms roughness. For the second derivative of the force, we have $\partial^2 F_N/\partial d^2 = F_N/l^2$. Substitution into (15) gives

$$W = 2\left(\pi^2 - 4\right)\frac{G^*}{E^*}\frac{F_N}{l^2}\left|u_z^{(0)}\right|\frac{v_x^{(0)2}}{\omega^2}. \tag{16}$$

Thus, for rough surfaces, the damping is proportional to the normal force and the amplitude of the normal oscillation. It can be tuned by changing the normal force, the oscillation amplitude and the oscillation frequency.

Finally, let us come back to the physical nature of the relaxation damping. Brillouin was probably the first to recognize that a non-vanishing dissipation at low velocity can only occur if there are some discontinuous jumps from one state to another in the system [25]. In other words, movement with finite velocity must occur in the system even if it is driven quasi-statically. Such rapid movements due to elastic instabilities are e.g. the reason for the appearance of finite dissipation in the celebrated Prandtl-Tomlinson-model [8]. At first glance, the oscillating contacts discussed in this Letter do not lead to any rapid movements. However, a singularity of stresses does exist at the border of the contact. This singularity leads to infinitely rapid movements even if the configuration of the contact changes quasi-statically. Let us illustrate this by the distribution of tangential stresses in the contact plane. The tangential stress distribution can be easily calculated from the linear force density $q_x(x)$ in the one-dimensional MDR-model by applying the integral transformation [11]

$$\tau(r) = -\frac{1}{\pi}\int_r^\infty \frac{q_x'(x)dx}{\sqrt{x^2 - r^2}}. \tag{17}$$

The tangential stress as a function of coordinate and time is shown in Fig. 4 as a color map. Of interest is the range of coordinates where the indenter is in contact only over some part of the oscillation period. In this range, one can see two maxima of the stress: the first one is located at the left boundary of the range. A detailed analysis shows that this is a logarithmic singularity, which is "pulsating" but not moving spatially. The second singularity is located at the right boundary of the contact; it develops and persists during the phase of the oscillation when the indenter is "pulled away". This is a "square root singularity", which is moving spatially. Movement of this singularity leads to infinitely rapid movements in the medium even if the indenter is moving quasi-statically.

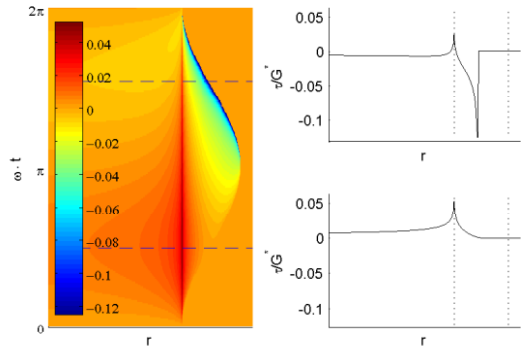

**Fig. 4.** Color map of the distribution of tangential stress as function of radius $r$ (horizontal axis) and time (vertical axis) over one period of the oscillation $z(t) = -z^{(0)} + u_z^{(0)} \cos \omega t$ and

$x(t) = x^{(0)} + u_x^{(0)} \cos(\omega t + \varphi_0)$ with the phase shift $\varphi_0 = \pi/2$.

At the beginning of the motion, a positive singularity appears at the initial boundary of the contact and remains at this point during the whole oscillation period (right lower subplot.) No energy dissipation is associated with this non-moving singularity. At the moment of reversal of the indentation movement (start of the "pulling" phase) a square-root-singularity appears at the right boundary of the contact and moves subsequently to the left, together with the shrinking contact region (right upper sub-plot.) At the same time, irreversible energy dissipation takes place. The right subplots correspond to the times shown in the color map with horizontal dashed lines. In the sub-plots, the maximum and the minimum extent of the contact region during an oscillation period is shown with dotted lines.

In conclusion, the effect of relaxation damping was discussed using the example of axis-symmetric elastic bodies with infinite friction in the contact area. The discussion was generalized to multi-contact systems and contact of bodies with rough surfaces. We have shown that a superposition of tangential movement (both with oscillating and constant velocity) and normal oscillations leads to a specific damping, which we call "relaxation damping". The damping is proportional to the amplitude of the normal oscillations and to the square of the tangential velocity. For nominally flat rough surfaces, it is also proportional to the applied normal force. The assumption of the infinite coefficient of friction was made only to study the effect in the pure. However, all results are directly applicable to systems with a finite coefficient of friction provided that the changes in the radius of the stick region are much smaller than those due to changing indentation. The effect will also be present, in a modified form, in sliding systems. Our analysis shows that application of normal oscillations will introduce additional damping of tangential movement into a system with friction. This may account for the well-known effect of suppression of frictional instabilities by application of ultrasonic oscillations, which was studied both theoretically [26] and experimentally [27].

We acknowledge useful discussions with N. Popov. This work was supported in part by the Ministry of Education of the Russian Federation, by COST Action MP1303 and by Deutsche Forschungsgemeinschaft.

[*] Corresponding author: m@popov.name

[1] Mindlin R.D., Mason W.P., Osmer J.F., Deresiewicz H. Effects of an oscillating tangential force on the contact surfaces of elastic spheres. Pros. 1st US National Congress of Applied Mechanics, ASME, New York, 203–208 (1952).

[2] Gaul L., Lenz J. Nonlinear dynamics of structures assembled by bolted joints. Acta Mechanica, 125 (4), 169-181 (1997).

[3] Akay A. Acoustics of friction. J. Acoust. Soc. of America, 111 (4), 1525-1548 (2002).

[4] Zhou X., Shin E., Wang K.W., Bakis C.E., Interfacial damping characteristics of carbon nanotube-based composites, Composites Science and Technology, 64 (15), 2425-2437 (2004).

[5] Davies M., Barber J.R., Hills D.A. Energy dissipation in a frictional incomplete contact with varying normal load. – Int. J. Mech. Sci.: 55, 13–21 (2012).

[6] Putignano C., Ciavarella M., Barber J.R. Frictional energy dissipation in contact of nominally flat rough surfaces under harmonically varying loads. J. of the Mechanics and Physics of Solids, 59, 2442–2454 (2011).

[7] Landau L.D., On the vibrations of the electronic plasma, Zh. Eksp. Teor. Fiz., 16, 574-586 (1946).

[8] Popov V.L. and Gray J.A.T., Prandtl-Tomlinson Model: History and applications in friction, plasticity, and nanotechnologies. - ZAMM, J. of Applied Math. Mech., 92, 683-708 (2012), DOI 10.1002/zamm.201200097.

[9] Meyer E., Overney R.M., Dransfeld K., Gyalog T., Nanoscience: Friction and Rheology on the Nanometer Scale (World Scientific, Singapore, 1998).

[10] Müser M.H., Urbakh M., Robbins M.O., Statistical mechanics of static and low-velocity kinetic friction, edited by I. Prigogine and S.A. Rice, Adv. Chem. Phys. 126, 187–272 (2003).

[11] Popov V.L., Heß M. Method of Dimensionality Reduction in Contact Mechanics and Friction, Springer, 2014, ISBN 978-3-642-53875-9.

[12] Galin L.A. Contact Problems in the Theory of Elasticity, 1961, North Carolina State College. This book is an English translation of the Russian original of 1953: Галин Л.А. Контактные задачи теории упругости. М., 1953.

[13] Sneddon I.N. The relation between load and penetration in the axisymmetric Boussinesq problem for a punch of arbitrary profile. Int. J. Eng. Sci., 3, 47-57 (1965).

[14] Cattaneo C. Sul contatto di due corpi elastici: distribuzione locale degli sforzi. Rendiconti dell'Accademia Nazionale dei Lincei 1938, 27: 342-348, 434-436, 474-478.

[15] Mindlin R.D. Compliance of elastic bodies in contact. Journal of Applied Mechanics., 16, 259–268 (1949).

[16] Jäger J. Axi-symmetric bodies of equal material in contact under torsion or shift. Archive of Applied Mechanics, 65, 478-487 (1995).

[17] Ciavarella M.: The generalized Cattaneo partial slip plane contact problem. I:Theory, International Journal of solids and structures, 35, 2349-2362 (1998).

[18] Heß M. Über die Abbildung ausgewählter dreidimensionaler Kontakte auf Systeme mit niedrigerer räumlicher Dimension. Göttingen: Cuvillier-Verlag, 2011.

[19] Munisamy R.L., Hills D.A., Nowell D. Static axisymmetrical Hertzian contacts subject to shearing forces. ASME J. Appl. Mech. 61, 278–283 (1994).

[20] Popov V.L. Contact mechanics and friction, Springer, 2010.

[21] Greenwood J.A. and Williamson J.B.P., Contact of Nominally Flat Surfaces. Proc. R. Soc. A 295, 300 (1966).

[22] Paggi M., Pohrt R.& Popov, V.L. Partial-slip frictional response of rough surfaces. Sci. Rep. 4, 5178 (2014); DOI:10.1038/srep05178.

[23] Pohrt R. and Li Q., Complete Boundary Element Formulation for Normal and Tangential Contact Problems, Physical Mesomechanics, 17 (4) (2014) in press.

[24] Campana C., Persson B.N.J., Müser M.H., J. Phys. Condens. Matter 23, 085001 (2011).

[25] Brillouin M., Notice sur les Travaux Scientifiques, Gauthiers-Vilars, Paris, 1990.

[26] Guerra R., Vanossi A., Urbakh M. Controlling microscopic friction through mechanical oscillations, Physical Review E, 78 (3), 036110 (2008).

[27] Popov V.L., Starcevic J. Effect of vibrations on the laboratory model "earthquake" statistics. Technical Physics Letters, 32, 630-633 (2006).